\documentclass[showpacs,amssymb,amsmath,nobibnotes, aps,prl,secnumarabic
,twocolumn%
]{revtex4-1}
\pdfoutput=1

\usepackage{bm}
\usepackage{natbib}
\usepackage{graphicx}

\begin{document}
\textfloatsep 10pt

\title{Inverse energy cascade and emergence of large coherent vortices in turbulence driven by Faraday waves}
\author{N. Francois}
\author{H. Xia}
\author{H. Punzmann}
\author{M. Shats}
\email{Michael.Shats@anu.edu.au}

\affiliation{Research School of
Physics and Engineering, The Australian National
University, Canberra ACT 0200, Australia}

\date{\today}

\begin{abstract}
We report the generation of large coherent vortices via inverse energy cascade in Faraday wave driven turbulence. The motion of floaters in the Faraday waves is three dimensional but its horizontal velocity fluctuations show unexpected similarity with two-dimensional turbulence. The inverse cascade is detected by measuring frequency spectra of the Lagrangian velocity, and it is confirmed by computing the third moment of the horizontal velocity fluctuations. This is observed in deep water in a broad range of wavelengths. The results broaden the scope of recent findings on Faraday waves in thin layers [A. von Kameke, \textit{et al.}, \textit{Phys. Rev. Lett.} \textbf{107}, 074502 (2011)].
\end{abstract}


\maketitle

Ripples appearing on the surface of a vertically vibrated liquid are known as Faraday waves. They are widely used to study a broad range of phenomena, such as pattern formation \cite{Kudrolli1996, Goldman2003, Kityk2005, Shani2010}, solitons \cite{Wu1984, Umbanhowar1986, Lioubashevski_1996, Xia2010, Rajchenbach2011, Shats2012}, walkers, or bouncing orbiting droplets \cite{Couder2005}, extreme wave events \cite{Shats2010}, and others. It has recently been found \cite{Kameke2011} that the horizontal motion of floating particles on the surface of Faraday waves for thin layers of liquids shows several properties consistent with the fluid motion in two-dimensional (2D) turbulence. This turbulent horizontal transport was referred to as Faraday flow. In particular, a Kolmogorov $k^{-5/3}$ spectrum has been reported as well as the Richardson law of pair dispersion \cite{Frisch}. Even though the layer was argued to be thin (i.e. smaller than the Faraday wavelength), the Faraday flows explored in these experiments are indisputably three dimensional (3D) and locally divergent flows. Some of the 2D turbulence theory assumptions are thus clearly violated and one should have few expectations for the emergence of the inverse energy cascade in this context.

However it has recently been observed in electromagnetically driven turbulence that the behavior of a 3D flow can be rendered similar to 2D turbulence in the presence of a large scale vortex \cite{Xia2011}. Moreover, recent numerical predictions emphasize that an inverse energy cascade might be a general principle existing in all turbulent flows in both 2D and 3D \cite{Biferale2012}. The emergence of such a phenomenology in Faraday waves would broaden the applicability of features common to 2D turbulent flows to surface wave phenomena. But the key ingredients supporting this proposition are to date still unclear. Indeed many features of Faraday flows are still unknown, such as the spectral localness of the energy injection in the surface ripples, the limitation to thin layers of fluid, or the dependence of the floaters motion on the frequency of vertical vibration. Besides it would be interesting to discover whether it is possible to generate large coherent structures, or spectral condensates, in the Faraday waves such as those observed in bounded electromagnetically driven turbulence \cite{Xia2009,Sommeria1986}.

In this paper we report new experimental results which show that the particle motion on the undulating surface is similar to the fluid motion in 2D turbulence. It supports the inverse energy cascade or the spectral energy transfer from smaller to larger scales. The water depth in these experiments is not restricted to thin layers, but rather substantially exceeds the Faraday wavelength. Moreover the vertical acceleration ranges from the Faraday instability threshold up to the droplet nucleation threshold where the ripples are a couple of millimeters high. Such a configuration rules out any 2D assumption on the fluid motion. If the flow is bounded, the inverse cascade leads to the accumulation of energy at the boundary box scale and the generation of a large vortex coherent over the whole domain.

\begin{figure}[t]
\includegraphics[width=7 cm]{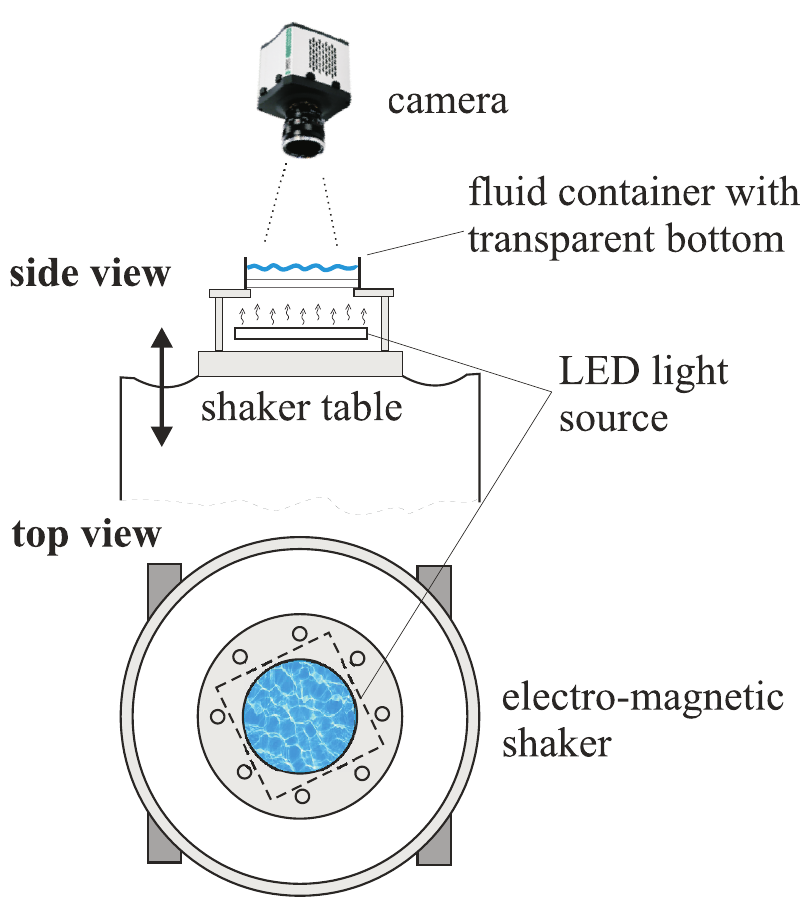}
\caption{\label{fig1} Experimental setup for the surface wave driven turbulence.}
\end{figure}

Experiments are carried out in a circular container (178 mm diameter, 30 mm deep) as shown in Figure~\ref{fig1}. The container is filled with a liquid whose depth is larger than the wavelength of the perturbations at the surface.  The monochromatic forcing is varied in the range of frequencies  $f_0=30-60$ Hz which results in Faraday wavelengths $\lambda$ in the range $\lambda\approx8-14$ mm. The dominant frequency of the excited surface ripples is at the first subharmonic of the excitation frequency, $f=f_0/2$.  We define the supercriticality as $\alpha=(a-a_{th})/{a_{th}}$ , where $a$ is the amplitude of the vertical acceleration imposed by the shaker and $a_{th}$ is the threshold for the parametric generation of Faraday waves.

\begin{figure}[t]
\includegraphics[width=8.75 cm]{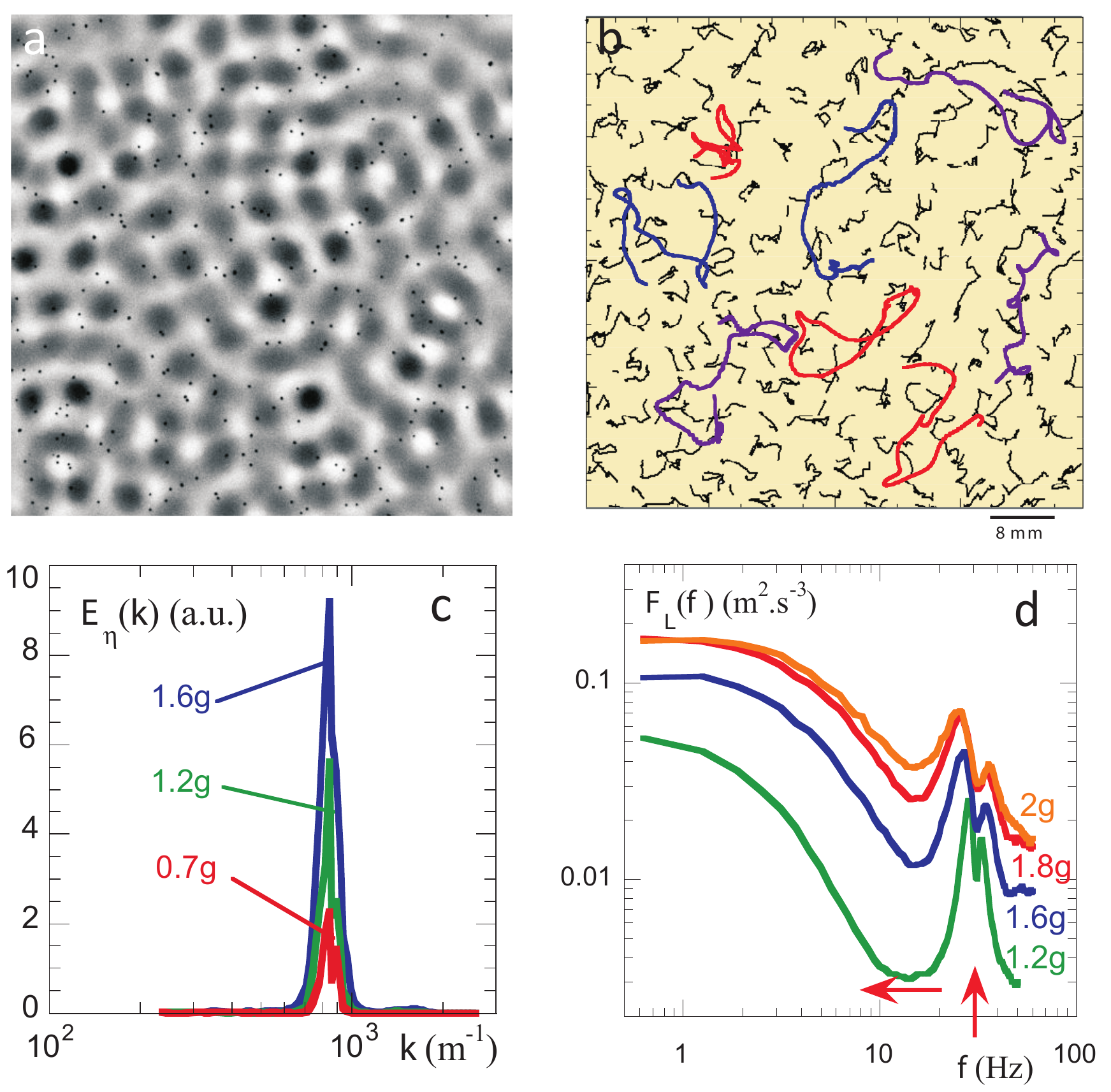}
\caption{\label{fig2} (a) Diffusive light image of the fluid surface elevation including floating particles. The particles diameter here is 300 microns \cite{particles}, the peaks and troughs of the oscillons appear as dark and white blobs. (b) Trajectories of oscillons (black) having a phase maximum, and trajectories of 8 particles (colored) tracked for 2 s (200 periods of the shaker). (c) Wave number spectra of the surface elevation at different vertical accelerations ($a=0.7$ g, 1.2 g, and 1.6 g) and $f_0=60Hz$. (d) Frequency power spectra of the Lagrangian velocity (averaged over $\approx1000$ trajectories) for different vertical accelerations at $f_0=60Hz$.}
\end{figure}

A diffusing light imaging technique is used to visualize simultaneously the surface ripples and floating tracer particles, as seen in Fig.~\ref{fig2}(a). A 2\% milk solution in water provides sufficient contrast for the parametrically excited waves to be observed and allows high resolution movies (800$\times$800 pixel) to be recorded in the range of speeds 80 - 120 fps using the Andor Neo sCMOS camera. Black floating particles are added on the fluid surface to visualize the horizontal motion of the liquid. Particles are made of carbon glass and have been plasma treated to reduce their intrinsic hydrophobicity. The use of surfactant and plasma treatment ensures that particles do not aggregate at the surface.  This is illustrated in Fig.~\ref{fig2}(a) with particles homogeneously distributed at the surface of the wave field.

It has recently been shown that the Faraday waves properties are consistent with those of an ensemble of bounded oscillating solitons, or oscillons \cite{Shats2012}. Figure~\ref{fig2}(b) shows examples of the oscillons' trajectories to be compared with a few particle tracks measured using the same tracking code. The horizontal motion of oscillons is random in time but confined to a disordered lattice as seen from the black trajectories shown in Fig.~\ref{fig2}(b) \cite{Xia2012}. The particles also wander erratically, but their trajectories show substantially larger excursions in comparison to oscillons. A particle usually visits several oscillon sites which randomize its trajectory. Although the particle and oscillons trajectories are characterized by completely different length scale, the oscillon mobility is an essential feature of Faraday flows. In the case of a perfect oscillonic crystal, in which the relative position of the nodes is frozen, the horizontal mobility of particles is zero, showing that the horizontal particle transport is strongly correlated to the oscillon motion. Such crystals are obtained by adding minute amounts of bovine serum to water \cite{Shats2012}.

Though the oscillon lattice in Fig.~\ref{fig2}(a) seems disordered, the wave number spectrum of the ripples is very narrow, as seen in Fig.~\ref{fig2}(c). This is consistent with the fact that oscillon motion is restricted in space to about half of the lattice characteristic wavelength \cite{Xia2012}. To generate 2D turbulence one needs to inject energy into the horizontal fluid motion at some intermediate range of scales in a localized wave number domain. The horizontal mobility of oscillons is the manifestation of a momentum transfer along the water surface; this transfer is restricted to a narrow \textit{k}-domain.

\begin{figure*}[!t]
\includegraphics[width=14 cm]{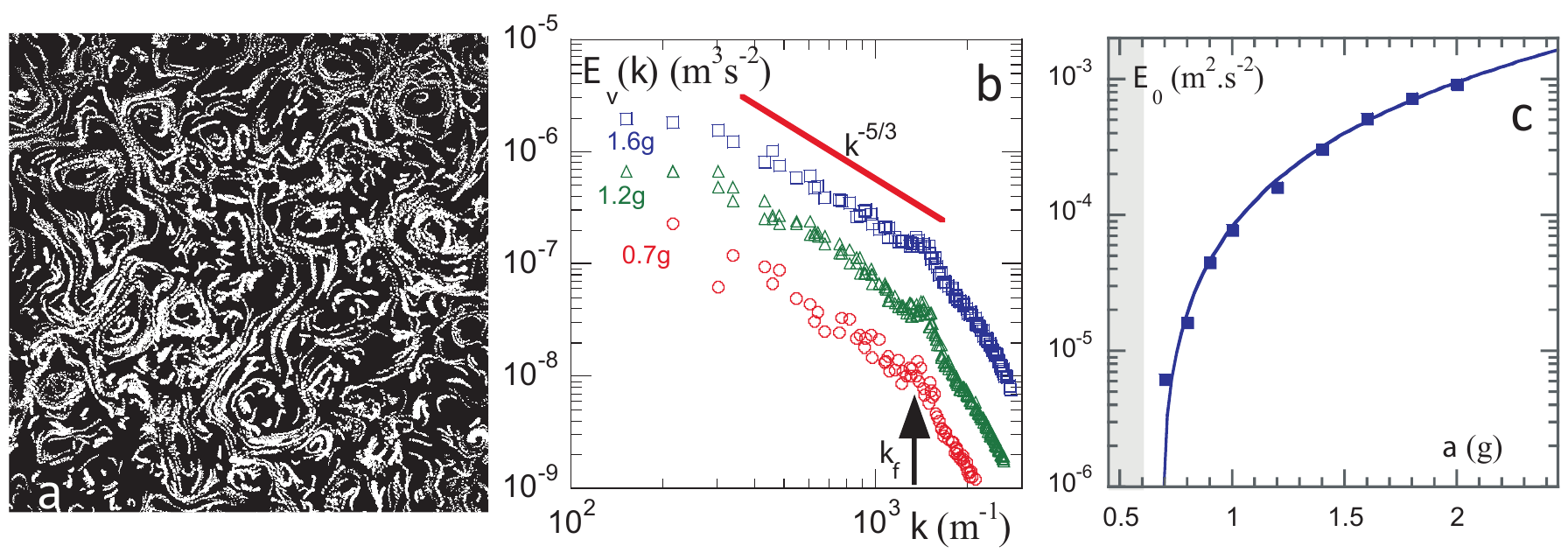}
\caption{\label{fig3} (a) Particle trajectories followed over 5 periods of the shaker oscillations at $a=1.6$g and $f_0=60Hz$ in a $8\times8$ cm$^{2}$ field of view. (b) Wave number spectra of the horizontal kinetic energy at different levels of the vertical acceleration ($a=0.7$ g, 1.2 g, and 1.6 g) and $f_0=60Hz$. (c) Kinetic energy of the flow $E_o$  versus vertical acceleration $a$. Grey bar indicates the threshold of parametric instability.}
\end{figure*}

Fig.~\ref{fig2}(d) shows frequency power spectra of velocity of particles moving along their Lagrangian trajectories. The spectra show peaks close to the Faraday frequency (30 Hz, the first subharmonic of the excitation frequency), and a broad low-frequency band whose energy grows with the increase in vertical acceleration. Since energy is injected at $f=30$ Hz, the growth and the broadening of the low-frequency band indicates a nonlinear energy transfer to slower temporal (larger spatial) scales. This transfer, as will be shown below, is due to the inverse energy cascade.

Fig.~\ref{fig3}(a) shows many trajectories of particles on a densely seeded fluid surface, the presence of multiple length scales in this Faraday flow is visible. In the following, Particle Image Velocimetry (PIV) is used to characterize the Eulerian velocity field \cite{technics}.

Figure~\ref{fig3}(b) shows wave number spectra of the horizontal kinetic energy measured at $f_0=60$ Hz at three vertical acceleration levels, $a=0.7$g, 1.2g, and 1.6g, corresponding to supercriticalities  in the range $\alpha=[0.17,1.7]$ ($a_{th}=0.6g$  at 60 Hz). The spectrum is close to the Kolmogorov-Kraichnan theory power law of $E_{\nu}(k)\propto k^{-5/3}$ at wave numbers $k \le 1500$ m$^{-1}$. At higher wave numbers, $k \geq 1500$ m$^{-1}$, spectra are somewhat steeper than expected from the direct enstrophy cascade fit of $k^{-3}$, probably due to higher damping at large wave numbers. If we define the turbulence forcing wave number from the position of the kink on the turbulence spectrum, e.g. $k_{f}\approx$1550 $m^{-1}$ at 60 Hz, Fig.~\ref{fig3}(b), it appears that the turbulence forcing wave number is roughly twice that of the surface elevation wave number (Fig.~\ref{fig2}(c)). The energy injection scale for the horizontal particle transport is thus related to the oscillon size which is about half of a period of the Faraday wave.

The kinetic energy of the horizontal flow can be estimated as $E_0=\int^{k_f}_{k_{low}} E_{\nu}(k) dk$. Here, $k_{low}$  is the lowest wave number determined by the field of view. In Fig.~\ref{fig3}(c), $E_0$ is shown as a function of the vertical acceleration $a$. By changing the vertical acceleration from 0.7g to 2.4g, we can vary the kinetic energy in the flow by over two orders of magnitude.

\begin{figure}[t]
\includegraphics[width=5.5 cm]{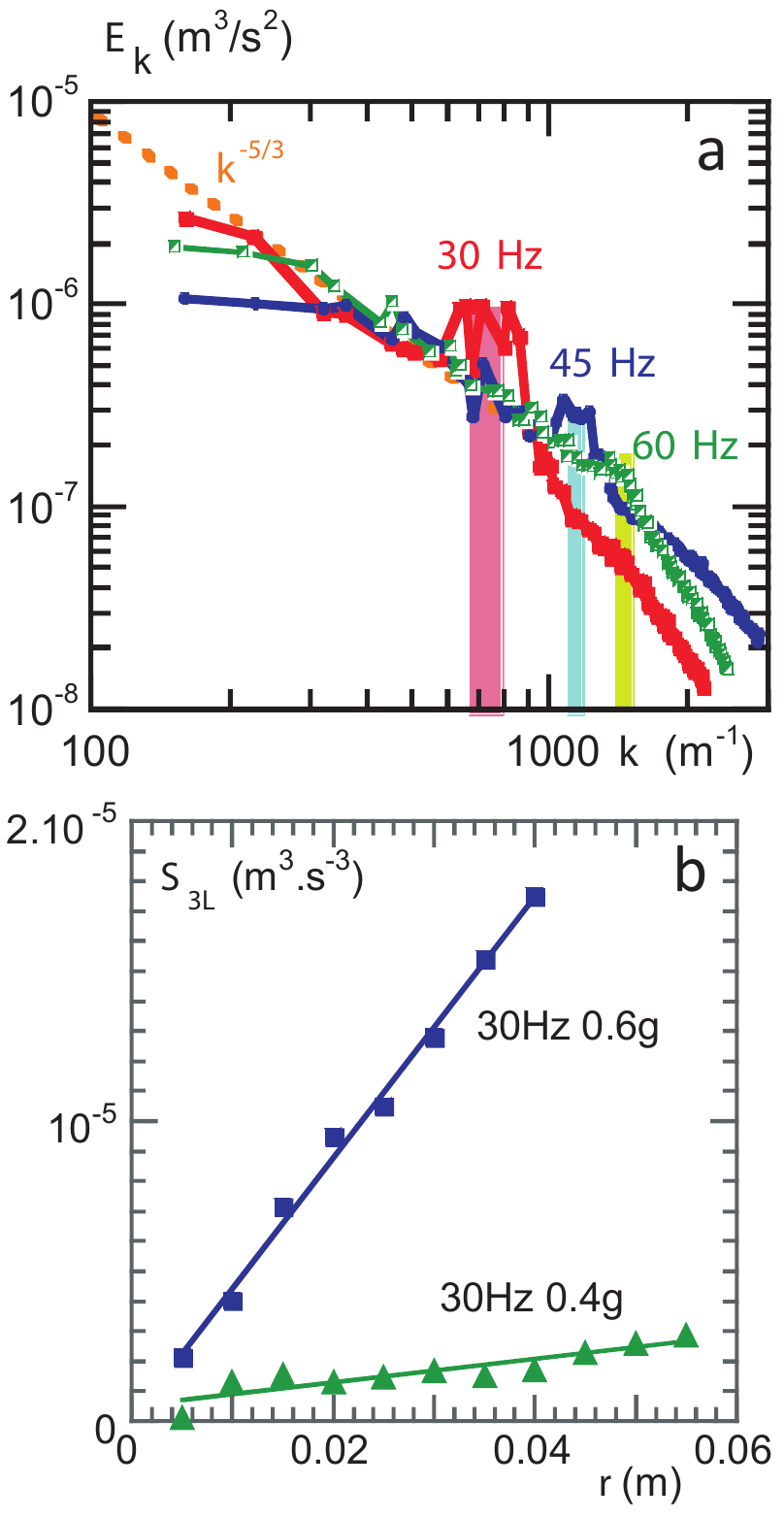}
\caption{\label{fig4}  (a) Energy spectrum in the surface wave driven turbulence at different frequencies (30 Hz, 45 Hz, 60 Hz), and at similar supercriticalities. Colored vertical bars show the forcing ranges. (b) Third moment of the longitudinal velocity increments measured at $f_0=30$ Hz at different supercriticalities  $\alpha=1 (a=$0.4g), and  $\alpha=2 (a=$0.6g).}
\end{figure}

Similar measurements were performed at several frequencies of the vertical vibrations: 30 Hz, 45 Hz, 60 Hz, Fig.~\ref{fig4}(a). A $k^{-5/3}$  power law is consistently observed in the energy spectra at lower wave numbers. A distinct kink is present in all spectra, and it is associated with the forcing wave number. The turbulence forcing wave number $k_f$ decreases with the decrease in frequency, in accordance with the capillary-gravity wave dispersion relation. A $k^{-5/3}$ range in the horizontal velocity spectra is consistent with the Kolmogorov-Kraichnan theory.

An important feature of 2D turbulence is the presence of the inverse energy cascade, or the transfer of energy from smaller to larger scales in the inertial range  $r_f < r < r_{low}$. A direct way to prove the existence of the inverse energy cascade is to compute the third moment of the velocity increments across a distance $r$ in a flow, $\delta v(r)=v(l+r)-v(l)$. The third order moment $S_{3L}=<[\delta v_L(r)]^3>$  (where index $L$ refers to the longitudinal velocity component with respect to $\textbf{r}$, and angular brackets denote averaging over all possible positions $l$ within the flow field, and in time, over many realizations) is related to the spectral energy flux $\varepsilon$  via the Kolmogorov flux relation. In the inertial range of 2D turbulence this relation reads as \cite{Xia2009,Yakhot1999}
\begin{equation}\label{eq2}
    S_{3L}=\frac{3}{2}\varepsilon r .
\end{equation}
Fig.~\ref{fig4}(b) shows third order structure functions computed from the velocity fields measured in the surface wave driven turbulence at $f_0=30Hz$ for two different vertical accelerations. In both cases $S_{3L}$ is a positive linear function of $r$. The spectral energy flux $\varepsilon$ can be computed from Eq.(1). The derived values are in agreement with the spectral power densities derived from the spectra of Fig.~\ref{fig4}(a).\\

\begin{figure}
\includegraphics[width=6.0 cm]{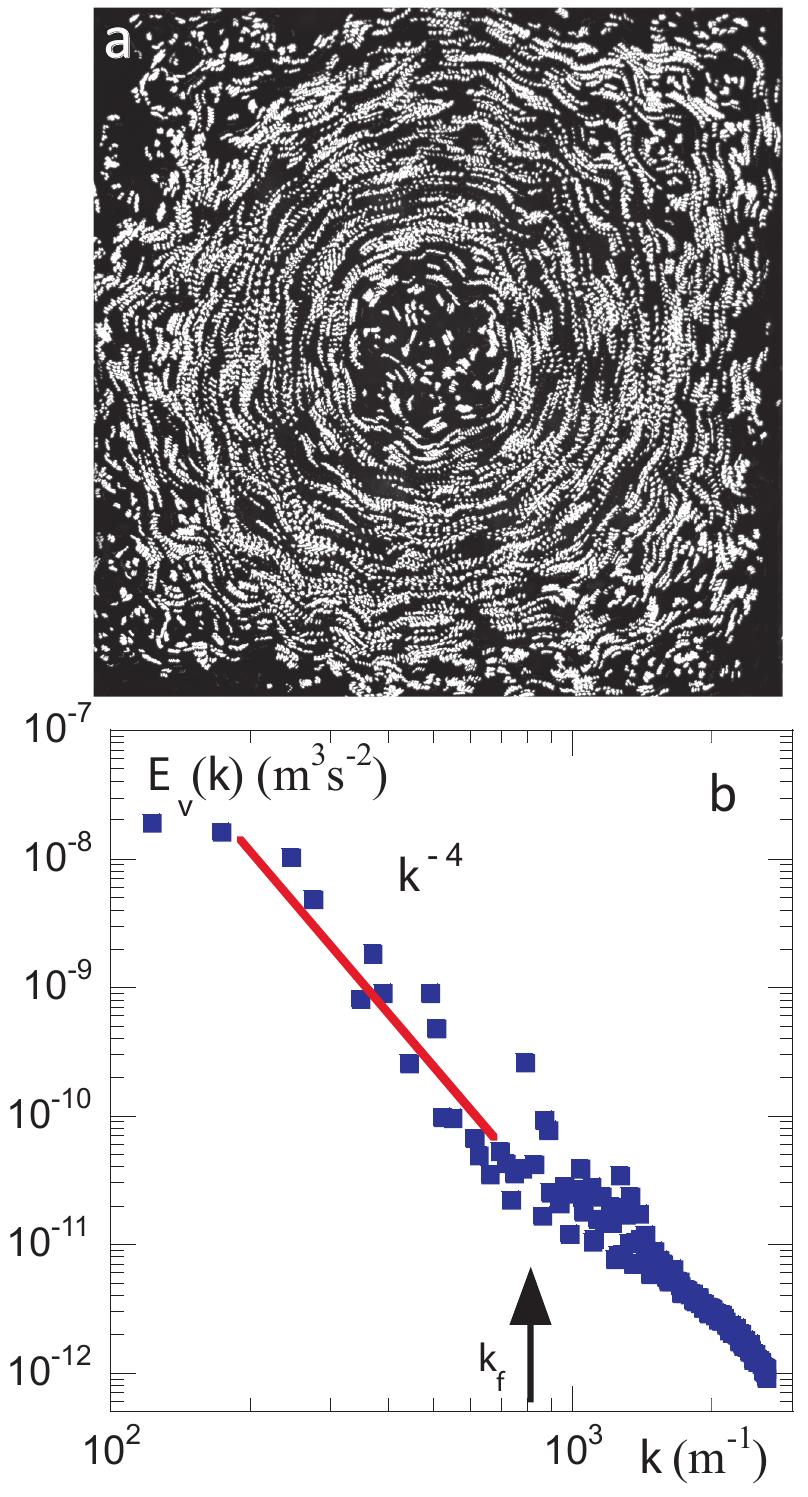}
\caption{\label{fig5} (a) Particles trajectories followed over 4 periods of the shaker oscillations in a spectrally condensed flow in a square box. The large scale vortex is 10 cm in diameter. The particles diameter is 50 microns. (b) Kinetic energy spectrum in the spectrally condensed turbulence.}
\end{figure}

The existence of the inverse energy cascade in a bounded flow is a prerequisite for the accumulation of the turbulence energy at the scale of the boundary size \cite{Kraichnan1967}. But the spectral condensation, or the generation of large coherent structures, is dependant on the boundary size, the energy flux, or the damping rate. It has been demonstrated in experiments in electromagnetically driven turbulence (EMT) that in a square box, whose size is smaller than the dissipation scale, turbulent inverse energy cascade can drive large scale vortices which are coherent over the entire domain \cite{Xia2009}. To test if the inverse energy cascade can generate coherent vortices in the surface wave driven turbulence, we produce Faraday flows in a square container ($110 \times 110$ mm$^2$) at the excitation frequency of $f_0\approx$ 30 Hz, which gives the forcing scale of $l_f \approx 7.7$ mm. Fig.~\ref{fig5}(a) shows trajectories of particles in this smaller square container. A large vortex occupying the entire box develops in the square container about 60 seconds after switching on the shaker. The kinetic energy spectrum of spectrally condensed flow is substantially steeper than the Kolmogorov spectra in Fig.~\ref{fig3}(b), $E(k) \propto k^{-4}$, as shown in Fig.~\ref{fig5}(b), similarly to the observations in EMT \cite{Xia2009}.\\

To conclude, experimental results support the existence of the inverse energy cascade fueled by Faraday waves. Faraday flows represent thus a new versatile tool of laboratory modeling of 2D turbulence. It is remarkable that the inverse energy cascade, initially thought of as a process intrinsic only to idealized 2D turbulence is found in a variety of physical systems where its existence could not possibly be expected \textit{a priori} \cite{Kameke2011,Xia2011,Byrne2011}. In 3D turbulence, the inverse energy cascade was recently found to naturally coexist with the direct energy cascade \cite{Biferale2012}.

\begin{acknowledgments}
This work was supported by the Australian Research Council's Discovery Projects funding scheme (DP110101525). HX would like to acknowledge the support by the Australian Research Council's Discovery Early Career Research Award (DE120100364).
\end{acknowledgments}


\begin{thebibliography}{02}
\bibitem{particles} For some runs, 50 $\mu$m diameter polyamide particles were used. No particle size effects were detected in comparison to runs involving 300 $\mu$m carbon glass particles.
\end{thebibliography}
\end{document}